\documentclass[a4paper]{jpconf}
\usepackage{graphicx}
\begin{document}

\title{Hydrodynamical evolution of dissipative QGP fluid}

\author{\bf A. K. Chaudhuri$^{1}$ and U. Heinz$^2$}
\address{$^1$Variable Energy Cyclotron Centre,
1/AF,Bidhan Nagar, Kolkata - 700 064\\
 $^2$Department of Physics,
The Ohio State University
Columbus, OH 43210}
\ead{akc@veccal.ernet.in; heinz@mps.ohio-state.edu}

\begin{abstract}
In the framework of the Mueller-Israel-Stewart theory of dissipative
fluid dynamics, we have studied the space-time evolution of a QGP fluid.
For simplicity, we have considered shear viscosity only and neglected
bulk viscosity and heat conductivity. Shear viscosity opposes the 
expansion and cooling of the  fluid. As a result, the lifetime of the 
fluid is extended. We also find that the parton $p_T$ distribution is
considerably flattened.
\end{abstract}

\section{Introduction}

One of the primary goals of relativistic heavy ion collisions is  the
creation and detection of the (lattice QCD \cite{lattice}) predicted 
deconfined phase of quarks and gluons. Experiments at RHIC, BNL, gave 
strong indications of QGP  formation in central Au+Au collisions
\cite{qm04}. Hydrodynamics  provides a simple, intuitive dynamic
description of relativistic heavy ion  collisions. Most of the
experimental data on soft hadron production at RHIC are well 
described  by ideal hydrodynamics. However, some problems remain. For 
example, one of the most important signals of collectivity, the elliptic
flow, saturates beyond $p_T\sim 1.5{-}2$~GeV/$c$ while ideal 
hydro\-dy\-namics predicts a continuous increase with $p_T$ \cite{he03}.

Ideal hydrodynamic may not give an accurate  description of the early
stage of the collision. In  the  early  stages  of collision, particle
momenta  are  predominantly  in the beam direction, while ideal 
hydrodynamic assumes a locally isotropic distribution.
The anisotropy in the momentum distribution will give rise to shear 
viscous stress. Other dissipative effects, e.g. bulk viscosity and 
heat conduction can also affect the hydrodynamic evolution of QGP. 
Unlike ideal flow, non-ideal flow generates entropy. The space-time 
evolution is also changed.

Relativistic theories for dissipative fluid dynamics were formulated 
long ago \cite{ec40,la59,2nd}. In early formulations, called 1st 
order theories \cite{ec40,la59}, the entropy 4-current contained terms 
of first order in the dissipative fluxes; these had the undesirable 
feature of violating causality. This is corrected in so-called 2nd 
order theories, due to Grad, Mueller and Israel and Stewart \cite{2nd}.
In 2nd order theories, the entropy 4-current contains terms of 2nd 
order in the dissipative fluxes. The space of thermodynamic variables 
is extended to include the dissipative fluxes, and relaxation equations  
for these dissipative fluxes are obtained from the positivity condition
for entropy production $\partial_\mu s^\mu \geq 0$.

Up to now, only a few authors have considered the effects of dissipation 
in relativistic heavy ion collisions \cite{ka84,da85,ch00,mu03,te03,te04,%
mu04}. Most of these studies considered one dimensional systems using 1st 
order theories. Only recently 2nd order theories have been implemented
numerically \cite{mu03,te04,mu04}. In the present paper, using the causal 
2nd order theory of dissipative fluid dynamics, we study the effect of 
shear viscosity on the space-time evolution of a QGP fluid. The paper is 
organized as follows: In Section II, we present the equations for 2nd order  
dissipative hydrodynamics. Our choice of equation of state, transport 
coefficients and initial conditions is described in Section III. 
Results for the space-time evolution of the QGP fluid, with and
without dissipation, are presented Section IV. Conclusions are drawn 
in Section VI.
 
\section{Equations for causal dissipative hydrodynamics}

We consider a QGP fluid in the central rapidity region, with zero net 
baryon density and chemical potential, $n_B=0,\ \mu_B=0$. We neglect 
the effects of heat conduction ($\mu_B=0$) and bulk viscosity (massless 
particles) and account only for shear viscosity. We work in the 
Landau-Lifshitz energy frame.

The energy-momentum tensor, including the shear viscous pressure 
tensor $\pi^{\mu\nu}$, is written as \cite{2nd}
\begin{equation}
T^{\mu\nu} = (\varepsilon+p)u^\mu u^\nu - p g^{\mu\nu} + \pi^{\mu\nu}
\end{equation}
where  $\varepsilon$ is the energy density, $p$ is the
hydrostatic pressure, and $u$ is the hydrodynamic 4-velocity,
normalized by $u^\mu u_\mu=1$. $T^{\mu\nu}$ satisfies the 
energy-momentum conservation law
\begin{equation}
\partial_\mu T^{\mu\nu} = 0.
\end{equation}

In the Israel-Stewart theory of dissipative fluids \cite{2nd}, the
dissipative fluxes are treated as independent thermodynamic variables
which satisfy kinetic relaxation equations. The transport (relaxation) 
equations for the shear viscous pressure read
\begin{equation}
\beta_2 D\pi^{\mu\nu}=-\frac{1}{2\eta} \pi^{\mu\nu} + 
\nabla^{\left\langle\mu\right.} u^{\left.\nu\right\rangle}
\end{equation}
where  $D=u^\mu\partial_\mu$ is the convective time derivative.  
$\eta$ is the shear viscosity coefficient. $\beta_2$ is related to 
relaxation time by $\tau_\pi=2\eta\beta_2$. The angular bracket in 
$\nabla^{\left\langle\mu\right.} u^{\left.\nu\right\rangle}$ is 
defined as
\begin{equation}
\nabla^{\left\langle\mu\right.} u^{\left.\nu\right\rangle}
  = \frac{1}{2}[ \nabla^\mu u^\nu+\nabla^\nu u^\mu] -\frac{1}{3}
    \Delta^{\mu\nu} \partial_\sigma u^\sigma,
\end{equation}
where $\nabla^\mu = \partial^\mu-u^\mu D$ is the transverse gradient
operator and $\Delta^{\mu\nu}=g^{\mu\nu}-u^\mu  u^\nu$ is the
projector orthogonal to the flow velocity $u^\mu$.

The viscous pressure tensor $\pi^{\mu\nu}$ is symmetric and traceless
$\pi^{\mu\nu} = \pi^{\nu\mu}$ and $\pi^\mu_\mu =0$. Furthermore, it is 
transverse to the hydrodynamic 4-velocity, $u_\mu \pi^{\mu\nu}=0$. Thus 
it has 5 independent components.

There are 10 unknowns ($\varepsilon$, $p$, three components of the 
hydrodynamic velocity  $u$, and 5 viscous pressure components) and 9 
equations (4 energy-momentum conservation equations and 5 transport 
equations for the independent components of $\pi^{\mu\nu}$. The system 
is closed by the equation of state $p=p(\varepsilon)$. We assume 
longitudinal boost-invariance and cylindrical symmetry in the transverse
direction. These additional symmetries reduce the independent variables
to four (energy density $\varepsilon$, radial velocity $v_r$ and
two components, say $\pi^{rr}$ and $\pi^{\phi\phi}$, of the viscous 
pressure tensor). In $(\tau,  r,  \phi,  \eta)$ coordinates, the
hydrodynamic and relaxation equations read (see appendix)
\begin{eqnarray} \label{5}
\partial_\tau \tilde{T}^{\tau \tau} 
+\partial_r \left(v_s \tilde{T}^{\tau \tau}\right)
 &=&- \frac{\tilde{p}-\tilde{\pi}^{rr}/\gamma^2  
    - r^2 \tilde{\pi}^{\phi\phi}}{\tau},\\
\partial_\tau \tilde{T}^{\tau r} 
+\partial_r\left(v_r \tilde{T}^{\tau r}\right)
&=&\frac{\tilde{p}+r^2\pi^{\phi\phi}}{r}
-\partial_r\left(\tilde{p} + \frac{\tilde{\pi}^{rr}}{\gamma^2}\right),\\
\partial_\tau \pi^{rr} +v_r \partial_r\pi^{rr}
&=&-\frac{1}{\tau_\pi \gamma}
\left(\pi^{rr} - 2\eta 
\nabla^{\left\langle r\right.}u^{\left.r\right\rangle}\right),\\
\partial_\tau \pi^{\phi\phi} +v_r \partial_r\pi^{\phi\phi}
&=&-\frac{1}{\tau_\pi \gamma} \left(\pi^{\phi\phi} 
   - 2\eta \nabla^{\left\langle\phi\right.}u^{\left.\phi\right\rangle}\right).
\end{eqnarray}
 
\noindent  where  $\tilde{A}^{\mu  \nu}  =  r  \tau A^{\mu \nu}$,
$\tilde{p}=r\tau p$ and $v_s=T^{\tau r}/T^{\tau\tau}$.
  
The four equations are solved simultaneously using the SHASTA-FCT
algorithm \cite{Shasta}. At each time step, we obtain the energy density
$\varepsilon$ and radial velocity $v_r$ from $T^{\tau\tau}$, $T^{\tau r}$,
and $\pi^{rr}$ by iteratively solving (in analogy to the ideal fluid 
case \cite{ri95})
\begin{eqnarray}
v_r = \frac{T^{\tau r}}{T^{\tau\tau}+p(\varepsilon)+\pi^{rr}/\gamma^2},
\qquad
\varepsilon = T^{\tau\tau} - v_r T^{\tau r}.
\end{eqnarray}

\begin{figure}[h]
\begin{minipage}{18pc}
\includegraphics[bb=13 80 581 740,width=18pc,clip=]{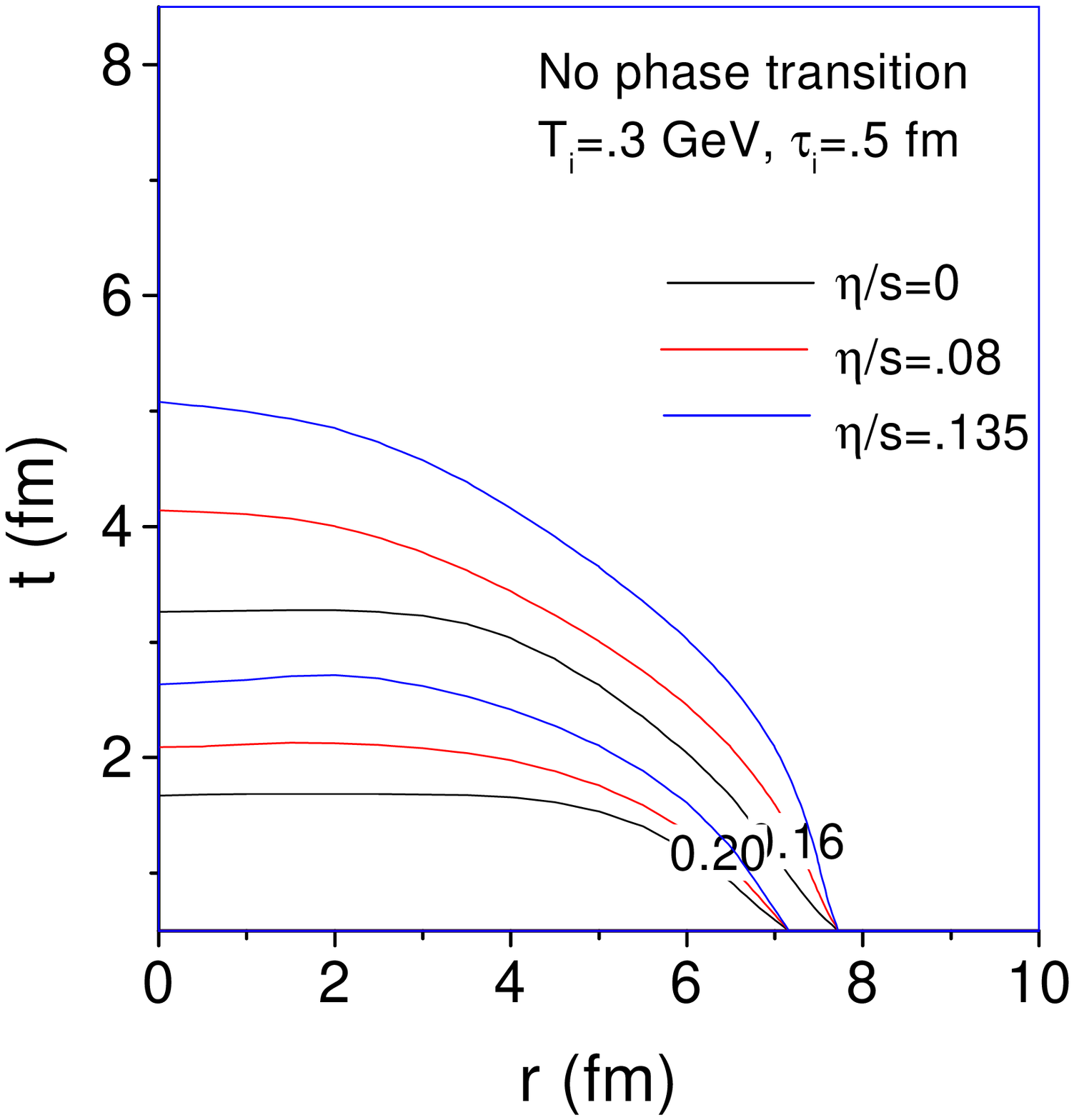}
\vspace{-2.5cm}
\caption{\label{F1}
Contours of constant temperature in the $r{-}\tau$ plane for ideal  
and non-ideal hydrodynamic evolution, using $T_i{\,=\,}0.3$~GeV at 
$\tau_i{\,=\,}0.5$\,fm/$c$.}
\end{minipage}\hspace{2pc}%
\begin{minipage}{18pc}
\includegraphics[bb=13 80 581 740,width=18pc,clip=]{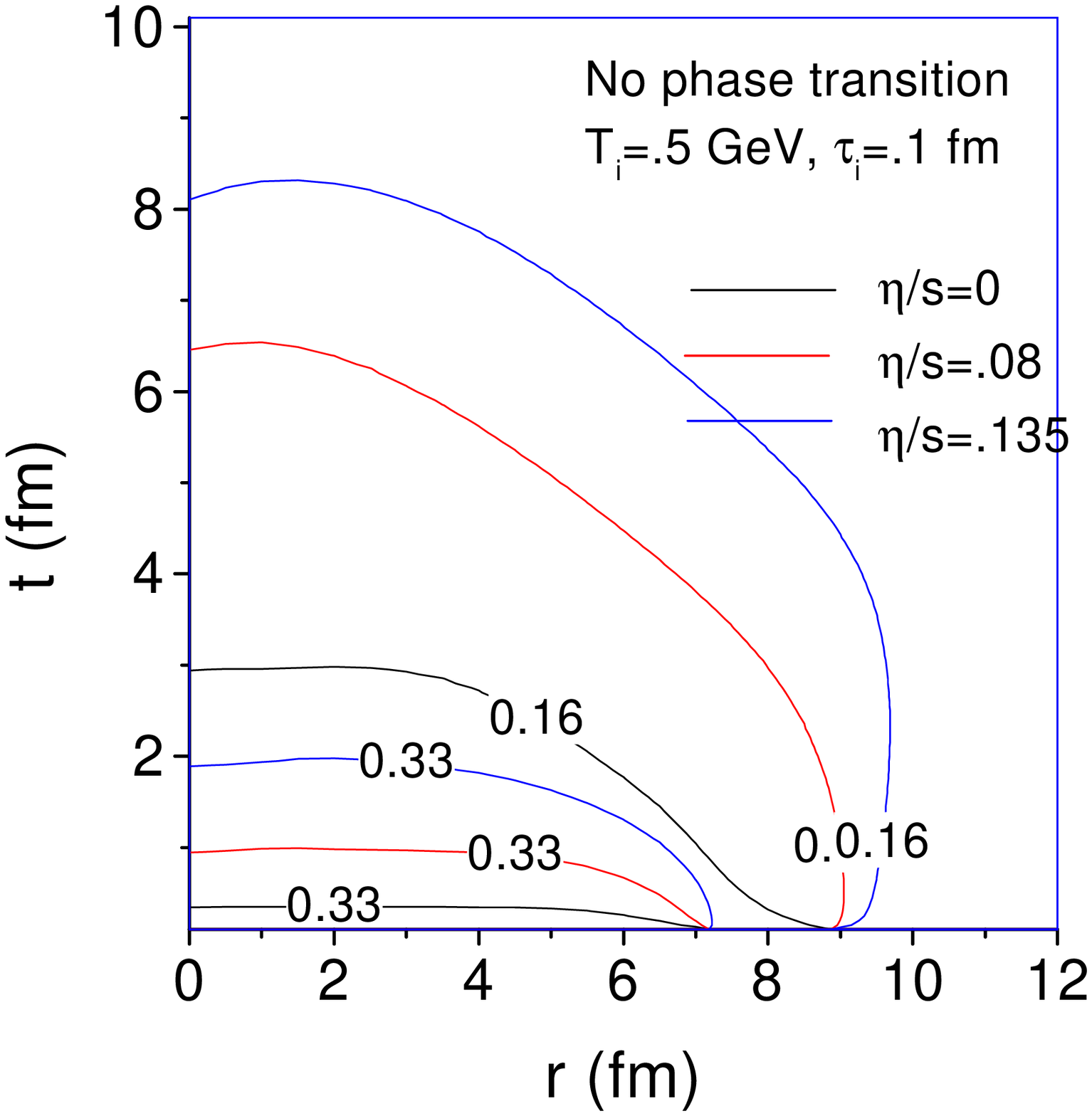}
\vspace{-2.5cm}
\caption{\label{F2}
Contours of constant temperature in the $r{-}\tau$ plane for ideal  
and non-ideal hydrodynamic evolution, using $T_i$=0.5 GeV at
$\tau_i{\,=\,}0.1$\,fm/$c$.}
\end{minipage} 
\end{figure}

\section{Equation of state, transport coefficient and initial conditions}

The equation of state is one of the important inputs for hydrodynamics.
It is the link between the macroscopic and microscopic world. In this 
exploratory work, we have used a simple equation of state,
$p=\frac{1}{3} \varepsilon$, with $\varepsilon=aT^4$, 
$a=(16+\frac{21}{2} N_f)\frac{\pi^2}{30}$.
 
In classical kinetic theory, explicit expressions can be obtained for the 
viscosity coefficient $\eta$ and the relaxation time $\tau_\pi$ in terms 
of the collision term. For a strongly coupled QGP, neither $\eta$ or 
$\tau_\pi$ are known. We treat them as phenomenological parameters. For 
guidance, we use perturbative \cite{ar00,ba90} and AdS/CFT \cite{po01}
estimates for $\eta$, respectively, and a kinetic theory estimate 
\cite{2nd} for $\tau_\pi$.

The shear viscosity coefficient $\eta$ for hot QCD was determined 
perturbatively to leading logarithmic accuracy in \cite{ba90,ar00}.
For $\alpha_s\approx0.5$ the result in \cite{ar00} gives
\begin{equation}
\frac{\eta}{s} = 0.135.
\end{equation}
Recently the shear viscosity was also evaluated in a strongly coupled 
gauge theory ($N=4$ SUSY YM theory), using the AdS/CFT correspondence  
\cite{po01}. The corresponding AdS/CFT estimate of shear viscosity is
\begin{equation} \label{14}
\frac{\eta}{s}=0.08
\end{equation}

In kinetic theory, in the Boltzmann gas approximation, the relaxation time
is estimated as $\tau_\pi=2 \eta \beta_2 =2\eta\, \frac{3}{4p}$ \cite{2nd}..

For  the initial energy density distribution in the transverse plane,
we use the Woods-Saxon parameterisation:
\begin{equation}
\varepsilon(r) =\frac{\varepsilon_0}{1+e^\frac{r-R}{a}},
\end{equation}
with  $R{\,=\,}6.4$~fm, $a{\,=\,}0.54$~fm. This is not very realistic,
but facilitates comparison with the results of \cite{mu04}. 
($\varepsilon_0{\,=\,}aT_i^4$ is the central energy density at 
initial time $\tau{\,=\,}\tau_i$.) We have considered two sets of 
initial conditions: (i) $T_i{\,=\,}0.3$~GeV and $\tau_i{\,=\,}0.5$~fm/$c$, 
and (ii) $T_i{\,=\,}0.5$~GeV and $\tau_i{\,=\,}0.1$~fm/$c$.
Both sets have similar total initial entropy content.

We also assume that at the initial time $\tau_i$ the radial velocity
vanishes ($v_r(r)=0$). 

For the non-ideal fluid, initial viscous pressures, $\pi^{rr}$ and
$\pi^{\phi\phi}$ are required. Even though $v_r$ and its derivatives 
are zero initially, due to the Bjorken longitudinal motion the stress 
tensor is not zero: $\nabla^{\left\langle r\right.} u^{\left.r\right\rangle>}
= r^2 \nabla^{\left\langle \phi\right.} u^{\left.\phi\right\rangle>}
=\frac{1}{3\tau_i}$. We assume that at initial time $\tau_i$, the
viscous pressure components are fully relaxed to the Bjorken scaling 
expansion values,
\begin{equation}
\pi^{rr}=r^2 \pi^{\phi\phi} = \frac{2\eta}{3\tau_i}.
\end{equation} 

\begin{figure}[h]
\begin{minipage}{12pc}
\includegraphics[bb=30 230 520 735,width=12pc,clip=]{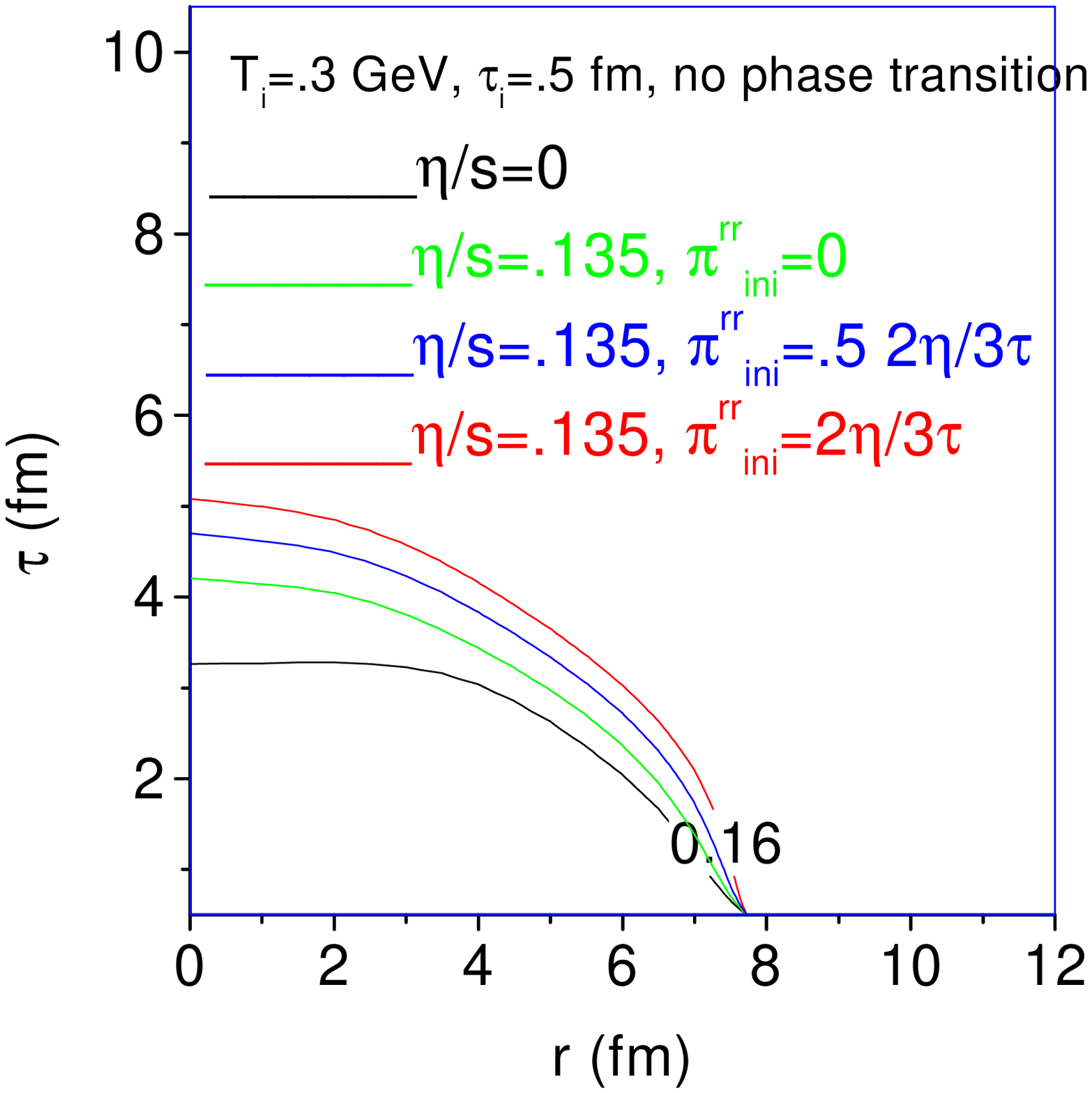}
\caption{\label{F3}
Freeze-out surface at $T_f{=}160$\,MeV for different initial viscous 
pressures.} 
\end{minipage}\hspace{1pc}%
\begin{minipage}{12pc}
\includegraphics[bb=30 230 520 735,width=12pc,clip=]{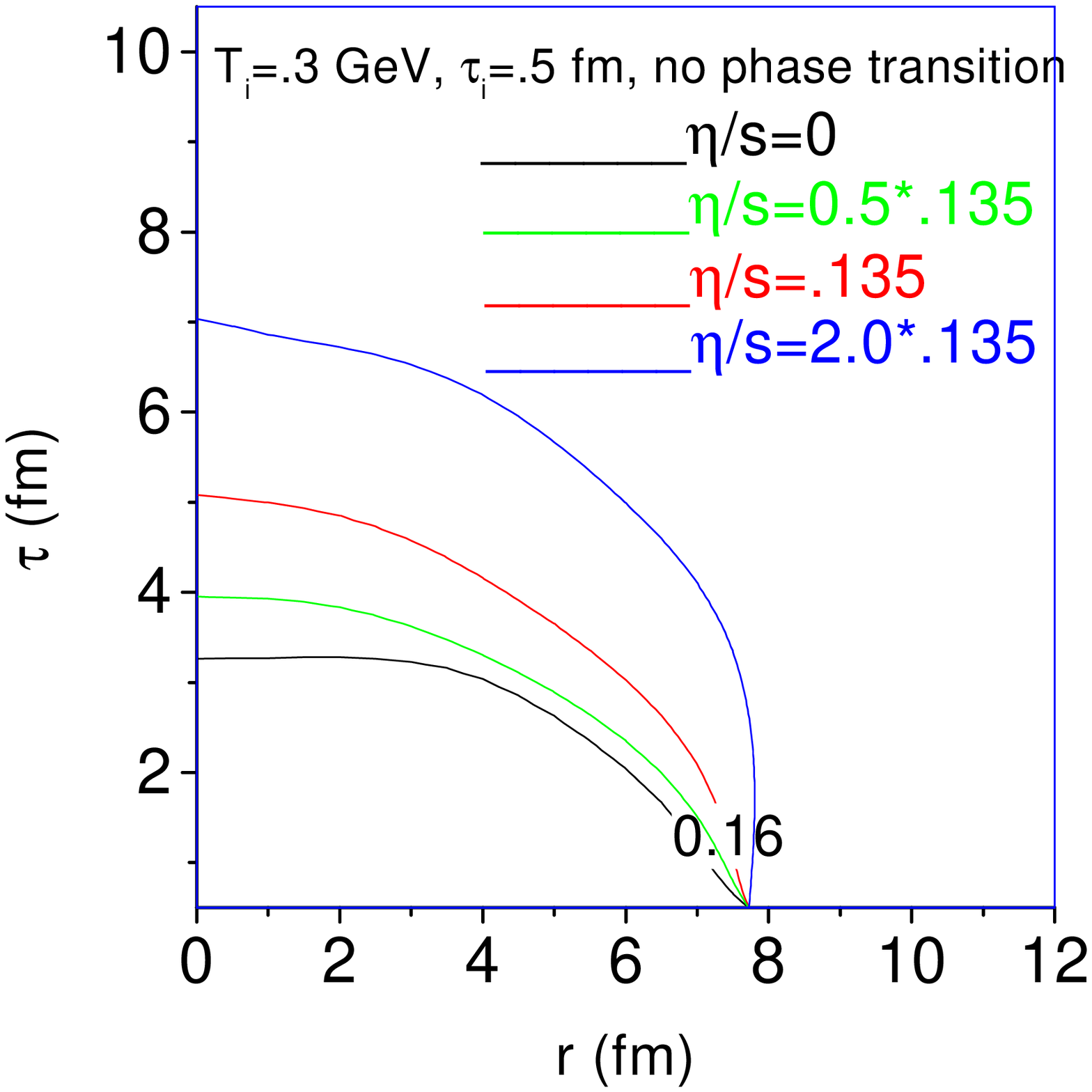}
\caption{\label{F4}
Freeze-out surface at $T_f{=}160$\,MeV for different shear viscosities
$\eta$.}
\end{minipage}\hspace{1pc}%
\begin{minipage}{12pc}
\includegraphics[bb=30 230 520 735,width=12pc,clip=]{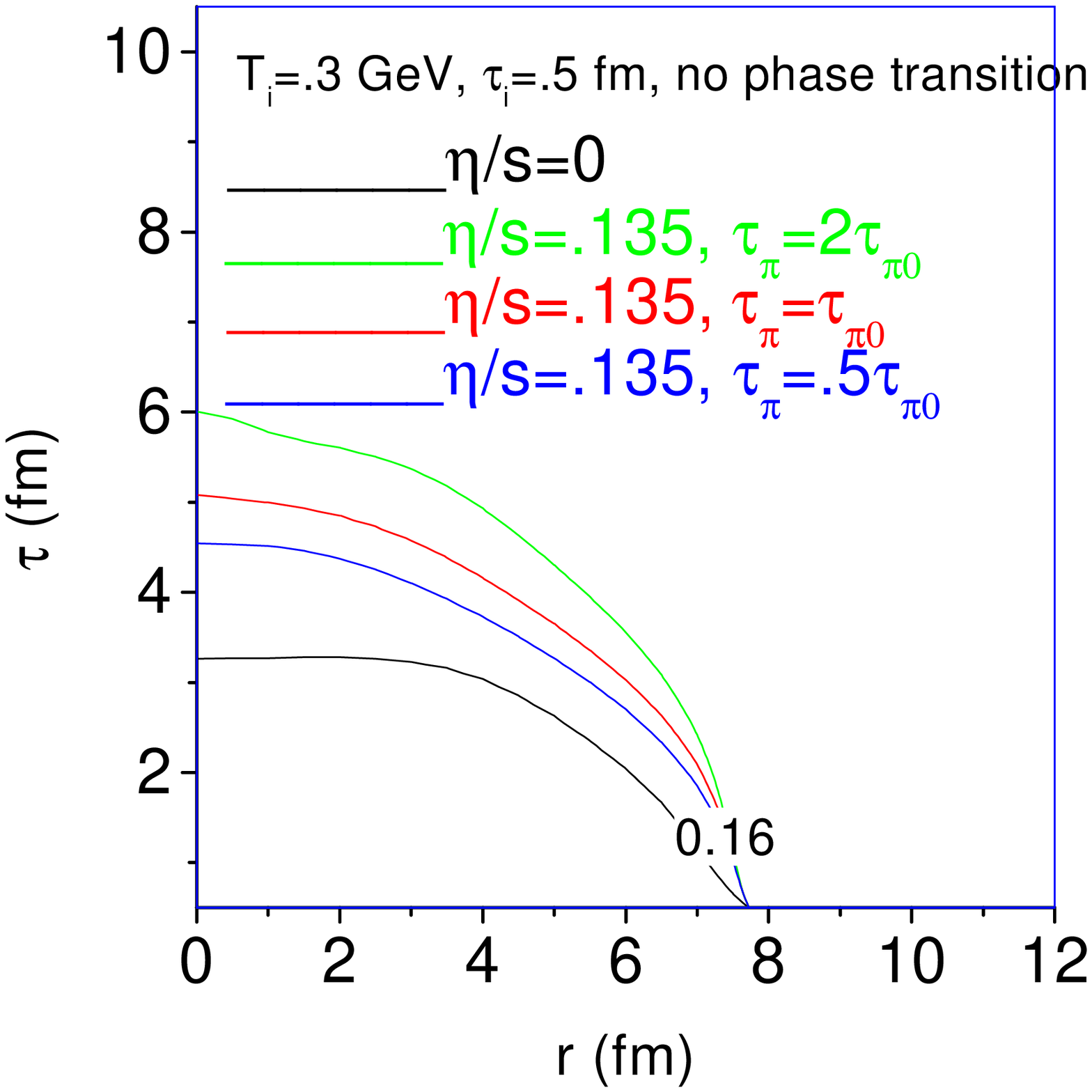}
\caption{\label{F5}
Freeze-out surface at $T_f{=}160$\,MeV for different relaxation times
$\tau_\pi$.}
\end{minipage} 
\end{figure} 

\section{Results}
\subsection{Space-time evolution of ideal and non-ideal QGP}

With have solved the hydrodynamic equation for ideal and non-ideal 
fluids, with identical initial conditions. We do not invoke a phase 
transition. The initial QGP evolves without undergoing phase transition 
until it freezes out at a an assumed freeze-out temperature 
$T_f{\,=\,}160$~MeV.

In Figs.~\ref{F1} and \ref{F2}, we show contours of constant temperature 
in the $r{-}\tau$ plane. The black lines are for the ideal fluid, while
the red and blue lines are for non-ideal fluids with $\eta/s$=0.08 and 
0.135, respectively. For ideal hydrodynamics the almost vanishing
slope of the isotherms near $r{\,=\,}0$ indicates that the transverse 
rarefaction wave has not reached the fireball center before the system
decouples at $T_f{\,=\,}0.16$~GeV. The cooling at $r{\,=\,}0$ is
controlled by the longitudinal scaling expansion. Consequently, the 
Bjorken cooling law, $T^3\tau$=constant, is well maintained at 
$r{\,=\,}0$. For non-ideal flow, the Bjorken cooling law is violated.
Viscosity opposes expansion and cooling. Thus the fluid cools more 
slowly and the lifetime of the fluid is extended. For a fluid with 
$T_i{\,=\,}0.3$~GeV at $\tau_i{\,=\,}0.5$~fm/$c$, the lifetime at $r{\,=\,}0$
is extended by 30\% and 55\% for $\eta/s$=0.08 and 0.135, respectively. 
For the higher initial temperature ($T_i{\,=\,}0.5$~GeV at 
$\tau_i{\,=\,}0.1$~fm/$c$), the viscosity effects are even more
prominent: in the fireball center the fluid life time is enhanced by 
more than 120\% and 170\% for $\eta/s$=0.08 and 0.135, respectively.
One also observes an increased transverse expansion.

The space-time evolution of non-ideal fluids depends sensitively on the
(i) initial viscous pressure, (ii) viscosity coefficient and (iii) relaxation
time. In Fig.~\ref{F3} the freeze-out surface at $T_f{\,=\,}0.16$~GeV
is shown for different initial viscous pressures,
$\pi^{rr}(=r^2\pi^{\phi\phi})$=0, $\frac{\eta}{3\tau_i}$ and
$\frac{4\eta}{3\tau_i}$, respectively, using $\eta/s$=0.135. The higher 
the initial viscous pressure, the more extended is the freeze-out surface. 
The life time of the dissipative QGP is extended by 20\% if the
initial viscous pressure is increased from zero to $\frac{4\eta}{3\tau_i}$.
The freeze-out surface also depends sensitively on the value of the 
viscosity coefficient (Fig.~\ref{F4}). As the viscosity decreases,
departure of the freeze-out surface from ideal behavior also 
decreases. In Fig.~\ref{F5} we show the freeze-out surface for different 
relaxation times, $\tau_\pi=0.5\, \tau_{\rm kin}$, $\tau_{\rm kin}$ and 
$2\,\tau_{\rm kin}$ (where $\tau_{\rm kin}=\frac{3\eta}{2p}$), for fixed
viscosity $\eta/s=0.135$. As the relaxation time is increased by a factor 
4, the freeze-out time in the fireball center is decreased by 25\%.

\begin{figure}[h]
\begin{minipage}{18pc}
\includegraphics[bb=30 230 520 790,width=18pc,clip=]{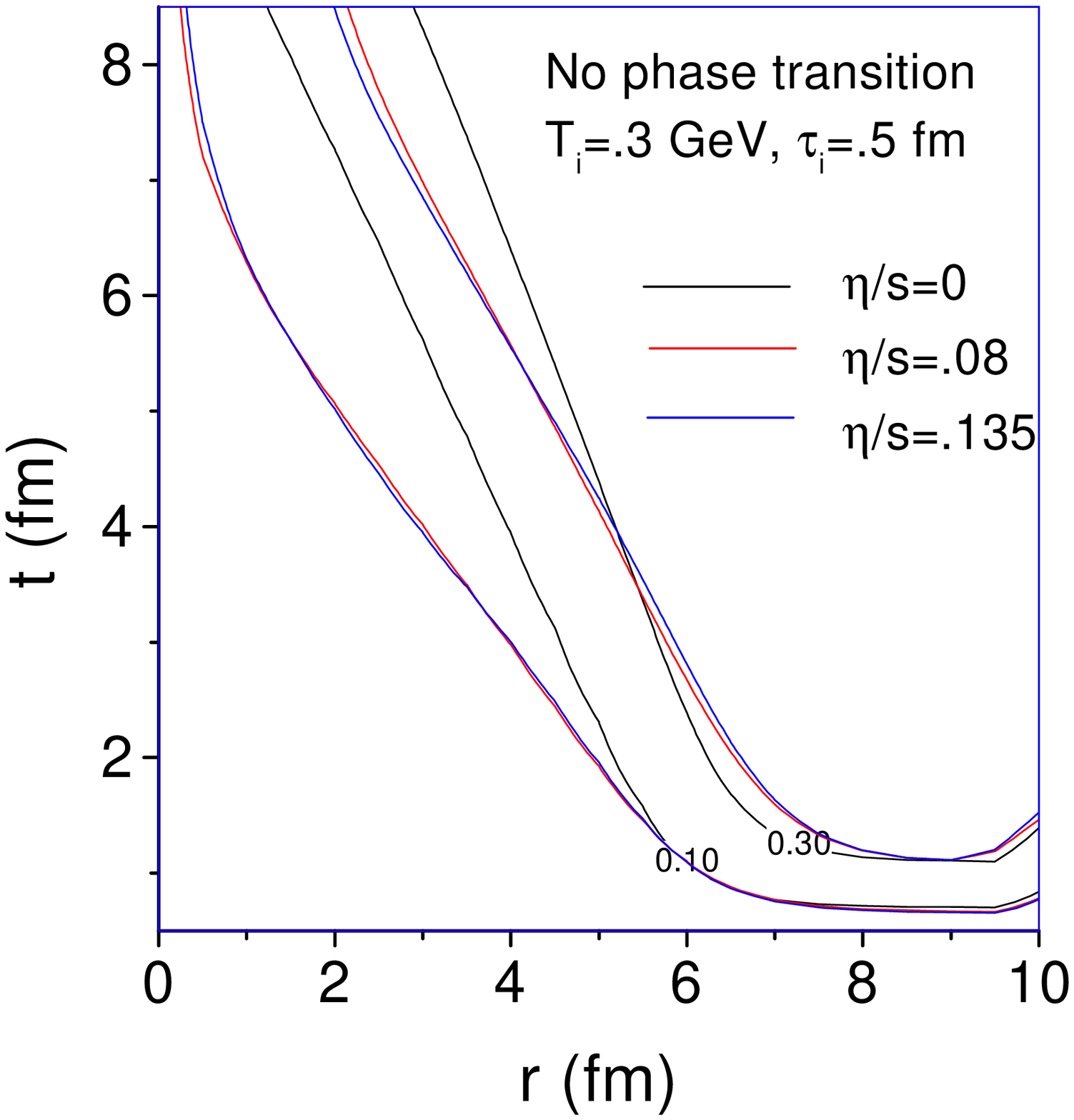}
\caption{\label{F6}
Contours of constant radial velocity in the $r{-}\tau$ plane for ideal  
and non-ideal hydrodynamic evolution, using $T_i$=0.3 GeV at 
$\tau_i$=0.5 fm/$c$.}
\end{minipage}\hspace{2pc}%
\begin{minipage}{18pc}
\vspace{0.5cm}
\includegraphics[bb=20 310 470 800,width=18pc,height=19pc,clip=]{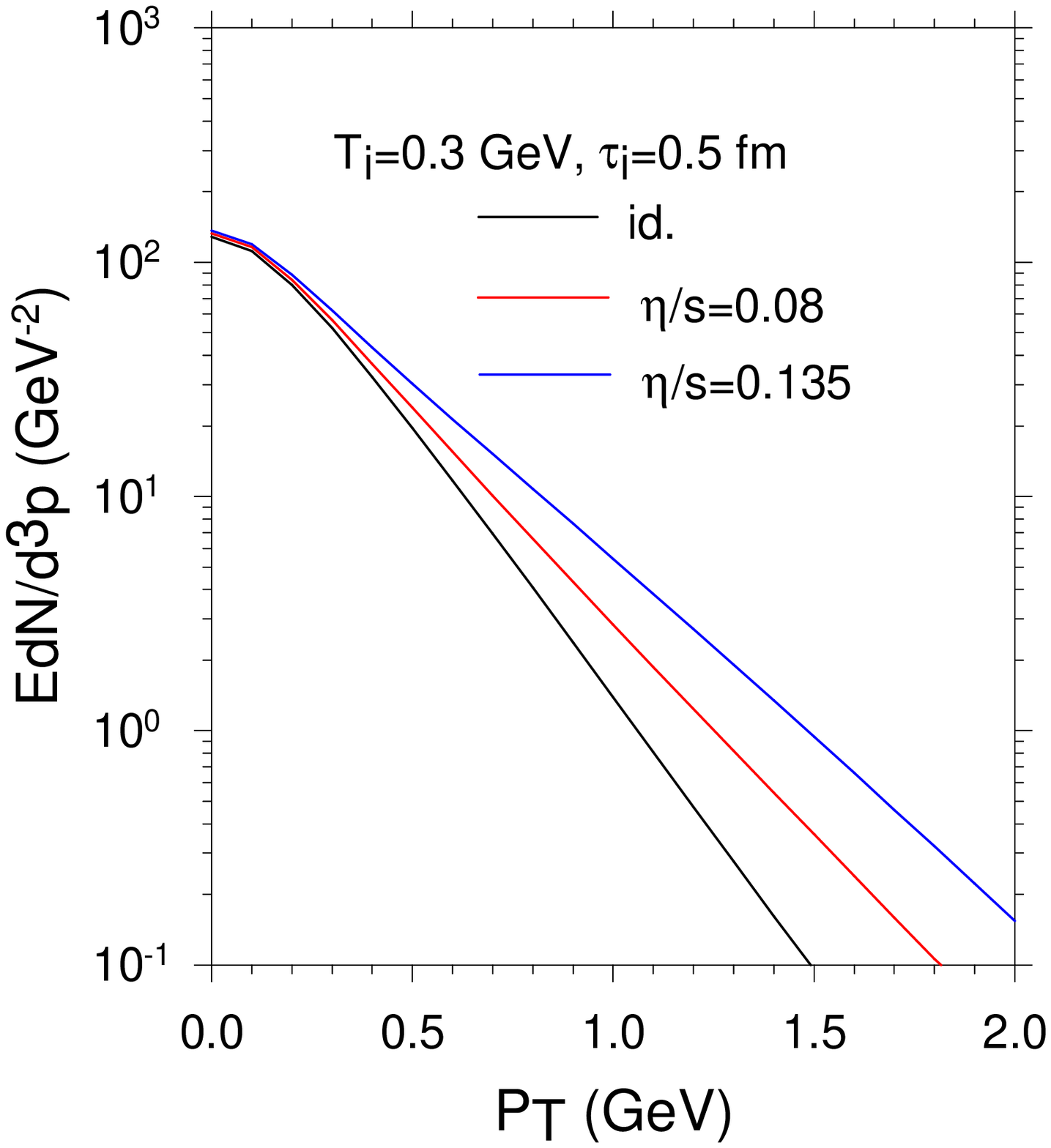}
\caption{\label{F7}
The final parton $p_T$ distribution at freeze-out ($T_f{\,=\,}160$\,MeV).
Initial temperature and time are $T_i{\,=\,}0.5$\,GeV and 
$\tau_i{\,=\,}0.1$\,fm/$c$.}
\end{minipage} 
\end{figure}

The evolution of the radial expansion velocity is also changed in viscous 
fluids. In Fig.~\ref{F6} we show lines of constant radial flow $v_r$ in 
the $r{-}\tau$ plane, for the ideal fluid (black line) and non-ideal 
fluids (red and blue lines). The radial velocity grow faster in the 
non-ideal fluid.

\subsection{Viscous corrections to the distribution function}

The changed space-time evolution of the dissipative QGP may have 
significant effects on particle production. Although we cannot
compute the $p_T$ spectra of final hadrons, since we did not include
the hadronization phase transition, we can calculate the $p_T$ spectra 
of partons on the $T{\,=\,}T_f$ freeze-out surface.

When viscous corrections are included, the distribution function is
modified from its local thermal equilibrium form \cite{2nd,te03,mu04}. 
When the viscous corrections are small, the distribution function can 
be written as
\begin{equation}
f(x,p)=f_{\rm eq}(x,p) (1 + \Delta f)
\end{equation}
with the equilibrium distribution function 
\begin{eqnarray*}
f_{\rm eq}(x,p) = \frac{g}{2\pi^2} \frac{1}{e^{u\cdot p/T} \pm 1}
\approx \frac{g}{2\pi^2} e^{-u\cdot p/T}.
\end{eqnarray*}
If $\Delta f$ is restricted to 2nd order in $p^\mu$, its form can
be determined as \cite{2nd,te03,mu04}
\begin{equation}
\Delta f = C p_\mu p_\nu \pi^{\mu\nu}
\end{equation}
where $C=\frac{1}{2(\varepsilon + p)T^2}$.

In Fig.~\ref{F7}, we show the parton $p_T$ spectra at $T_f{\,=\,}160$~MeV. 
The black line is for the ideal  fluid.  The red and blue lines are for 
non-ideal fluids with $\eta/s$=0.08 and 0.135, respectively. Viscous flow 
is seen to enhance high $p_T$ production by a factor 5--10, depending on
viscosity. However, since $\pi^{\mu\nu}\neq0$ already at initial time 
$\tau_i$, already the initial $p_T$ spectra for the viscous fluids are 
flatter than for the ideal fluid. To obtain the same measured final 
$p_T$-spectrum in the viscous case thus requires a retuning of initial 
conditions. This is presently under investigation.
 
\section{Summary and conclusions} 

We have studied the effects of shear viscosity on the space-time evolution 
of a relativistic QGP fluid. In the 2nd order theory of dissipative fluid
dynamics, the dissipative fluxes are treated as thermodynamic variables 
which follow kinetic relaxation equations. These are solved simultaneously 
with the energy-momentum conservation equations. Viscosity opposes
expansion and cooling. Consequently, for a fixed freeze-out temperature, 
the freeze-out time (fluid lifetime) is increased. The evolution of 
the radial flow velocity is also changed -- the viscous fluid generates more
transverse flow. Viscous effects flatten the $p_T$ distribution and 
require a retuning of the initial conditions if the final spectra
are to be kept fixed at their measured values. 

{\bf Acknowledgement:} The work of U.H. was supported by the U.S.
Department of Energy under contract DE-FG02-01ER41190.

\section*{References} 


\appendix
\section{Coordinate transformations}

Instead of Cartesian coordinates $x^\mu=(t,x,y,z)$ we use curvilinear
cylindrical coordinates in longitudinal proper time and rapidity,
$\bar{x}^m=(\tau,r,\phi,\eta)$:
\begin{eqnarray}
t &=& \tau \cosh\eta ; \hspace{1.6cm} \tau=\sqrt{t^2-z^2} \\
x &=& r \cos\phi ; \hspace{1.6cm} r=\sqrt{x^2+y^2}\\
y &=& r \sin\phi; \hspace{1.6cm} \phi=\tan^{-1}\left(\frac{y}{x}\right)\\
z &=& \tau \sinh\eta; \hspace{1.6cm} \eta=\frac{1}{2}\ln \frac{t+z}{t-z}.
\end{eqnarray}
The differentials 
\begin{eqnarray}
dt &=& d\tau \cosh\eta + d\eta\, \tau \sinh\eta, \\
dx &=& dr\, \cos\phi - d\phi\, r \sin\phi, \\
dy &=& dr\, \sin\phi + d\phi\, r \cos\phi, \\
dz &=& d\tau \sinh\eta+ d\eta\, \tau \cosh\eta,
\end{eqnarray}
and the metric tensor is easily read off from
\begin{eqnarray}
  ds^2 &=& g_{\mu\nu} dx^\mu dx^\nu = dt^2-dx^2-dy^2-dz^2 
\nonumber\\
       &=& \bar g_{mn} d\bar x^m d\bar x^n =
           d\tau^2 -dr^2 -r^2 d\phi^2 -\tau^2 d\eta^2,
\end{eqnarray}
namely
\begin{equation}
\bar g_{mn}=\left(
\begin{array}{cccc}
   1      & 0   & 0   & 0   \\
   0      & -1  & 0   &  0  \\
   0      & 0   &  -r^2  & 0   \\
   0      &  0  & 0  &  -\tau^2  \\
\end{array} \right), \hspace{1cm}
\bar g^{mn}=\left(
\begin{array}{cccc}
   1      & 0   & 0   & 0   \\
   0      & -1  & 0   &  0  \\
   0      & 0   &  -1/r^2  & 0   \\
   0      &  0  & 0  &  -1/\tau^2  \\
\end{array} \right)
\end{equation}
In curvilinear coordinates we must replace the partial derivatives 
with respect to $x^\mu$ by covariant derivatives (denoted by a 
semicolon) with respect to $\bar x^m$:
\begin{eqnarray*}
{\bar T^{ik}}_{;p} &=&\frac{\partial \bar T^{ik}}{\partial \bar x^p} 
+ \Gamma^i_{pm} \bar T^{mk} + \bar T^{im}\Gamma^k_{mp}.
\end{eqnarray*}
The only non-vanishing Christoffel symbols are
\begin{equation}
\Gamma^\tau_{\eta\eta}=\tau;
\quad
\Gamma^\eta_{\tau\eta}=\Gamma^\eta_{\eta\tau}=1/\tau;
\quad
\Gamma^r_{\phi\phi}=-r;
\quad
\Gamma^\phi_{r\phi}=\Gamma^\phi_{\phi r}=1/r.
\end{equation}
The hydrodynamic 4-velocity $u^\mu=\gamma(1,v_x,v_y,v_z)$ is transformed 
to $\bar u^m=\gamma(1,v_r,0,0)$, with $\gamma=1/\sqrt{1{-}v^2_r}$.
From here on, we drop the bars over tensor components in 
$\bar x$-coordinates for simplicity.

\section{Relaxation equations for the viscous pressure tensor}

Being symmetric and traceless, the viscous pressure tensor $\pi^{\mu\nu}$
has 9 independent components. The assumption of boost invariance
reduces this number by 3 
($\nabla^{\left\langle m\right.}u^{\left.\eta\right\rangle}{\,=\,}0, \ 
m \neq \eta$). The additional assumption of cylindrical symmetry 
eliminates two more components 
($\nabla^{\left\langle m\right.}u^{\left.\phi\right\rangle}{\,=\,}0, 
\ \eta\neq{m}\neq\phi$. The transversality condition $u_m \pi^{mn}=0$
eliminates another two components ($u_\phi$ and $u_\eta$ vanish and 
thus yield no constraints). Thus, with boost-invariance and cylindrical 
symmetry, the viscous pressure tensor has only two independent components 
which we here choose as $\pi^{rr}$ and $\pi^{\phi\phi}$.

The viscous pressure tensor relaxes on a short kinetic time scale $\tau_\pi$
to $2\eta$ times the shear tensor $\sigma^{\mu\nu}=
\nabla^{\left\langle\mu\right.}u^{\left.\nu\right\rangle}$ \cite{2nd}.
The relaxation equations for $\pi^{rr}$ and $\pi^{\phi\phi}$ are
\begin{eqnarray}
\partial_\tau \pi^{rr}+v_r \partial_r \pi^{rr} &=&
-\frac{1}{\tau_\pi \gamma}\left(\pi^{rr} - 2\eta 
\nabla^{\left\langle r\right.}u^{\left.r\right\rangle}\right),\\
\partial_\tau \pi^{\phi\phi}+v_r \partial_r \pi^{\phi\phi} &=&
-\frac{1}{\tau_\pi \gamma}\left(\pi^{\phi\phi} - 2\eta 
\nabla^{\left\langle\mu\right.}u^{\left.\phi\right\rangle}\right).
\end{eqnarray}
The $(rr)$ and $(\phi\phi)$ components of the shear tensor can be written as
%
\begin{eqnarray}
\nabla^{\left\langle r\right.}u^{\left.r\right\rangle}
&=&-\partial_r u^r -u^r Du^r -\frac{1}{3} \Delta^{rr} 
\partial{\cdot}u = \gamma^2 \left(\frac{\partial{\cdot}u}{3}
-\partial_\tau\gamma -\partial_r(\gamma v_r)\right),\\
\nabla^{\left\langle\phi\right.}u^{\left.\phi\right\rangle}
&=&-\frac{1}{r^2} \frac{u^r}{r} -\frac{1}{3} \Delta^{\phi\phi} 
\partial{\cdot}u = \frac{1}{r^2}\left(\frac{\partial{\cdot}u}{3}
-\frac{\gamma v_r}{r}\right),
\end{eqnarray}
with the scalar expansion rate
\begin{equation}
\partial{\cdot}u 
= \partial_\tau u^\tau + \partial_r u^r+ \frac{u^\tau}{\tau} +\frac{u^r}{r}
= \left(\partial_\tau+\frac{1}{\tau}\right)\gamma 
+ \left(\partial_r   +\frac{1}{r}\right)(\gamma v_r)
\end{equation}
and the convective detivative
\begin{equation}
D = u\cdot\partial = \gamma(\partial_\tau + v_r \partial_r).
\end{equation}
The projectors $\Delta^{rr}$ and $\Delta^{\phi\phi}$ are,
\begin{eqnarray*}
\Delta^{rr}&&=g^{rr}- u^r u^r = -1 -(u^r)^2 = -\gamma^2\\
\Delta^{\phi\phi}&&=g^{\phi\phi}- u^\phi u^\phi
=-\frac{1}{r^2}.
\end{eqnarray*}
The remaining non-vanishing components of the viscous pressure tensor
$\pi^{\tau  \tau}$, $\pi^{\tau  r}$, and $\pi^{\eta\eta}$ are eliminated
using the constraints $u_m \pi^{mn} =0$ and $g_{mn} \pi^{mn}=0$, yielding
\begin{eqnarray}
\pi^{\tau\tau} = v_r \pi^{\tau r} =  v^2_r \pi^{rr},
\quad
\pi^{\eta \eta} = -\frac{1}{\tau^2} \left(-\frac{\pi^{rr}}{\gamma^2}
 +r^2 \pi^{\phi\phi}\right).
\end{eqnarray}

\section{Energy-momentum conservation}

With longitudinal boost-invariance and cylindrical symmetry, the 
energy-momentum conservation equations ${T^{mn}}_{;n}=0$ yield 
\begin{eqnarray} 
\partial_\tau \tilde{T}^{\tau \tau} 
+\partial_r\left(v_s \tilde{T}^{\tau \tau}\right)
&=& -\frac{\tilde{p}-\tilde{\pi}^{rr}/\gamma^2 
    - r^2 \tilde{\pi}^{\phi\phi}}{\tau},\\
\partial_\tau \tilde{T}^{\tau r} 
+\partial_r \left(v_r \tilde{T}^{\tau r}\right)
&=&\frac{\tilde{p}+r^2\pi^{\phi\phi}}{r}
-\partial_r\left(\tilde{p} + \frac{\tilde{\pi}^{rr}}{\gamma^2}\right).
\end{eqnarray}
where $\tilde{A}^{mn}\equiv  r\tau A^{mn}$,
$\tilde{p}\equiv r\tau p$, and $v_s \equiv T^{\tau r}/T^{\tau\tau}$.

\end{document}